
\documentclass[twocolumn]{aastex631}
\usepackage{amsmath}
\usepackage{graphicx}

\begin{document}


\title{Rotation of Polarization Angle in Gamma-Ray Burst Prompt Phase. III. The Influence of the Magnetic Field Orientation}

\author{Xing-Yao Wang}
\author{Jia-Sheng Li}
\author[0000-0001-5641-2598]{Mi-Xiang Lan}
\affiliation{Center for Theoretical Physics and College of Physics, Jilin University, Changchun, 130012, China; lanmixiang@jlu.edu.cn}

\begin{abstract}
Polarization is very sensitive to the configuration of the magnetic field in the radiation region. In addition to polarization curve and polarization spectrum, studies of polarization angle (PA) rotation spectrum is also crucial. In this paper, we use a simple parametric magnetic reconnection model with a large-scale aligned magnetic field in the radiation region to study the effects of field orientation on the PA rotations. Under different field orientations, variations of the PA rotation with parameters and the PA rotation spectra are studied. We find that the conclusions obtained in our previous works are almost independent of the field orientations. The area of the parameter space with $\Delta$PA $>10^\circ$ will shrink as the value of field orientation ($\delta$) increases for $0^\circ<\delta<90^\circ$. The $\Delta$PA values would be the same for two complementary field orientations. For two particular magnetic field orientations ($\delta=0^\circ$ and $90^\circ$), the $\Delta$PA would also only be $0^\circ$ or $90^\circ$ within the burst duration.
\end{abstract}


\shorttitle{Rotation of Polarization Angle}
\shortauthors{Wang, Li, \& Lan}

\section{Introduction}

Gamma-ray bursts (GRBs) are the most violent burst events occurring in the Universe \citep{2004RvMP...76.1143P, 2006RPPh...69.2259M, 2015PhR...561....1K}. It is generally believed that long bursts originate from the collapse of massive stallar cores and short bursts from the mergers of two neutron stars \citep{1992ApJ...395L..83N, 1993ApJ...405..273W, 1999Natur.401..453B, 2003Natur.423..847H, 2017ApJ...848L..13A, 2018PhRvL.120x1103L}. Currently, there are three major models to explain the observations of the GRB prompt emission: the internal shock model \citep{1994ApJ...427..708P, 1994ApJ...430L..93R}, the photospheric model \citep{1994MNRAS.270..480T, 2000ApJ...529..146E, 2005ApJ...628..847R, 2013MNRAS.428.2430L}, and the magnetic reconnection model \citep{2011ApJ...726...90Z}. Polarization predictions of the three models are different \citep{2020ApJ...892..141L, 2021ApJ...909..184L, 2024ApJ...970...10L, 2025ApJ...979..219L, 2020ApJ...896..139P, 2022ApJ...926..104P}. Therefore, polarization detection is particularly important in order to distinguish between these three models.

Since polarization is sensitive to the magnetic field configuration of the radiation region, polarization studies can constrain the magnetic field properties \citep{2003ApJ...594L..83G, 2009ApJ...698.1042T, 2019ApJ...870...96L}. In previous studies, \cite{2020ApJ...892..141L} investigated the time- and energy-resolved polarization of the GRB prompt phase. \cite{2024MNRAS.529.4287S} and \cite{2021RAA....21...55L} used the $q-\Pi$ curves to discuss the statistics of the polarizations in GRB prompt phase. \cite{2024ApJ...970...10L} and \cite{2025ApJ...979..219L} studied the multiwavelength polarization curves and spectra ranging from optical band to MeV. These studies provided systematic polarization predictions in GRB prompt phase.

The study of the polarization angle (PA) is as important as that of the polarization degree (PD). Previous studies have shown that the PA evolves differently for the aligned and toroidal field configurations during the early afterglow phase \citep{2016ApJ...816...73L}. \cite{2023ApJ...946...12W} found that the PA shows rotations for slight off-axis observations. \cite{2024ApJ...973....2L} then identified three parameters that affect PA rotation significantly for the aligned field. \cite{2024A&A...687A.128C} discussed the influence of the similar parameters on the PA rotation for the toroidal field. The PA rotation spectra from optical to MeV gamma-rays of both toroidal and aligned fields have been studied by \cite{2025ApJ...979..219L}. However, all of these studies have neglected the effect of the key parameter,  the orientation of the aligned field on polarizations. As expected the orientation of the aligned field does not effect the predicted PD values, while its impact on the PA rotation is rather significant. Therefore, the influence of the orientation of the aligned field on the PA rotations should be investigated.

In this paper, we use a simple parametric magnetic reconnection model to calculate the PA rotations during the GRB prompt phase. The effects of the orientations of the aligned field on the PA rotations are studied. This paper is arranged as follows: In Section \ref{sec:The Model} we briefly introduce the model we use, in Section \ref{sec:Numerical Results} we show our numerical results, and in Section \ref{sec:Conclusions and Discussion} the conclusions and discussion are given.

\section{The model}\label{sec:The Model}

As in the previous studies \citep{2015ApJ...808...33U, 2016ApJ...825...97U, 2018ApJ...869..100U}, the radiation region is assumed to be a relativistic thin shell expanding radially from the central engine at redshift $z$. The electrons in the shell are accelerated via the magnetic reconnection process to emit synchrotron photons in the magnetic field. The shell is assumed to begin emitting at radius $r_{on}$ and stop at radius $r_{off}$. The total number of radiating electrons $N$ in the shell at $r_{on}$ is $0$ and increases at an injection rate $R'_{inj}$.

Here, we use a three-segment power-laws photon spectrum $H_{en}(\nu')$ \citep{2024ApJ...970...10L, 2025ApJ...979..219L}:
\begin{equation}
H_{en}(\nu')=
\begin{cases}
(\nu'/\nu'_1)^{\alpha_1+1}, & \text{$\nu'<\nu'_1$}, \\
(\nu'/\nu'_1)^{\alpha_2+1}, & \text{$\nu'_1< \nu'<\nu'_2$}, \\
(\nu'_2/\nu'_1)^{\alpha_2+1}(\nu'/\nu'_2)^{\beta+1}, & \text{$\nu'>\nu'_2$},
\end{cases}
\end{equation}
where $\nu'=\nu(1+z)/\mathcal{D}$, $\nu'_1=\min(\nu'_{cool}, \nu'_{min})$ and $\nu'_2=\max(\nu'_{cool}, \nu'_{min})$. Here, $\nu$ is the observational frequency and $\mathcal{D}=1/[\Gamma(1-\beta\cos\theta)]$ is the Doppler factor. In addition, $\alpha_1$, $\alpha_2$ and $\beta$ are the low-energy, mid-energy, and high-energy photon spectral indices, respectively. In this paper, we take $\alpha_1=-2/3$, $\alpha_2=-1-(p-1)/2$ and $\beta=-1-p/2$ in a slow-cooling phase ($\nu'_{cool}>\nu'_{min}$) \citep{1979rpa..book.....R, 1998ApJ...497L..17S}, where $p$ is the index of the true energy spectrum ($N(\gamma_e)\propto\gamma_e^{-p}$) of the injected electrons. And we take $\alpha_1=-2/3$, $\alpha_2=-1$ and $\beta=-1-p/2$ in the fast-cooling phase ($\nu'_{cool}<\nu'_{min}$) \citep{2014NatPh..10..351U, 2016A&A...588A.135Y, 2021ApJ...913...60P}.

Here, the two characteristic frequencies read $\nu'_{cool}=3q_eB'\gamma_{cool}^2\sin\theta'_{B'}/4\pi m_ec$ and $\nu'_{min}=3q_eB'\gamma_{ch}^2\sin\theta'_{B'}/4\pi m_ec$, respectively. And $q_e$ is electron charge, $m_e$ is electron rest mass, $B'$ is the magnetic field strength in the co-moving frame, $\theta'_{B'}$ is the pitch angle between the direction of the magnetic field and the electron's velocity, and $c$ is the speed of light. The cooling Lorentz factor of the electrons can be expressed as:
\begin{equation}
\gamma_{cool}=\frac{6\pi m_ec\Gamma}{\sigma_TB'^2t},
\end{equation}
where $t$ is the dynamical time in the burst-source frame and $\sigma_T$ is the Thomson cross section. And for $\gamma_{ch}$ we use the model $i$ proposed by \cite{2018ApJ...869..100U}, which is expressed as
\begin{equation}
\gamma_{ch}(r)=\gamma_{ch}^0\biggl(\frac{r}{r_0}\biggr)^g,
\label{gammach}
\end{equation}
where $\gamma^0_{ch}$ is the normalization value of $\gamma_{ch}$ at radius $r_0$, and $r_0$ is the normalization radius. We take $g=-0.2$ for i model \citep{2018ApJ...869..100U}.

The emitting shell is also accelerated and the variation of its bulk Lorentz factor $\Gamma$ is a power law with the radius $r$ from the central engine \citep{2002A&A...387..714D}
\begin{equation}
\Gamma(r)=\Gamma_0\biggl(\frac{r}{r_0}\biggr)^s,
\end{equation}
where $\Gamma_0$ is the normalization value of $\Gamma$ at radius $r_0$, and we take bulk acceleration index $s=0.35$ for an aligned field in the shell \citep{2001A&A...369..694S}. In the co-moving system, the magnetic field strength $B'$ can be formulated as
\begin{equation}
B'(r)=B'_0\biggl(\frac{r}{r_0}\biggr)^{-b},
\label{B'}
\end{equation}
where $B'_0$ is the normalization value of $B'$ at radius $r_0$, and $b$ is the magnetic field decay index.

The polarization model used here is the same as that in \cite{2020ApJ...892..141L} and \cite{2024ApJ...970...10L}. The detailed calculation formula can be found there. In this paper, we only consider the case of a large-scale ordered aligned magnetic field, which corresponds to the magnetic reconnection process in the ejecta of a pulsar whose rotational axis is perpendicular to its magnetic axis \citep{2001A&A...369..694S}. The time- and energy-resolved polarization degree (PD, $\Pi$) and preliminary PA ($\text{PA}_\text{pre}$, $\chi_\text{pre}$) of the radiation from a jet with an aligned field in its emission region can be expressed as follows.
\begin{equation}\label{Pi}
\Pi=\frac{\sqrt{Q_{\nu}^2+U_{\nu}^2}}{F_{\nu}},
\end{equation}
\begin{equation}\label{chi}
\chi_\text{pre}=\frac{1}{2}\arctan\biggl(\frac{U_{\nu}}{Q_{\nu}}\biggr),
\end{equation}
where $F_{\nu}$ is the observed flux density, $Q_{\nu}$ and $U_{\nu}$ are the Stokes parameters, respectively. However, the observed PA can't be obtained only by the above formula, and the sign of the Stokes parameters should be considered to get the real PA. When $Q_\nu>0$, then the final PA value is $\chi=\chi_\text{pre}$. When $Q_\nu<0$, if $U_\nu>0$ then the final PA value is $\chi=\chi_\text{pre}+\pi/2$ and if $U_\nu<0$ is $\chi=\chi_\text{pre}-\pi/2$ \citep{2018ApJ...860...44L}.

\section{Numerical results}\label{sec:Numerical Results}

In this section, we apply the model described in Section \ref{sec:The Model} to discuss the impact of the field orientation on the PA rotations. Here, we define $\Delta \text{PA}=\text{PA}_\text{max}-\text{PA}_\text{min}$ to denote the rotation of the PA within the $T_{90}$ range ($T_{90}=T_{95}-T_5$), $T_5$ and $T_{95}$ are the times when the time-cumulated flux reach $5\%$ and $95\%$ of the total time-integrated flux, respectively, where $\text{PA}_\text{max}$ and $\text{PA}_\text{min}$ are the maximum and minimum values of PA in the time range from $T_5$ to $T_{95}$, respectively. In our calculations, unless otherwise specified, the parameters are set as follows: the jet opening angle $\theta_j=0.1$ rad, $\gamma_{ch}^0=5\times10^4$, $r_0=10^{15}$ cm, $r_{on}=10^{14}$ cm, $r_{off}=3\times10^{16}$ cm, $\Gamma_0=250$, $B'_0=30$ G, $b=1$ \citep{2018ApJ...869..100U, 2018A&A...609A.112G, 2019MNRAS.488.5823L, 2023ApJ...959...13R}, and $p=2.6$.

\begin{figure*}
\centering
\includegraphics[scale=0.35]{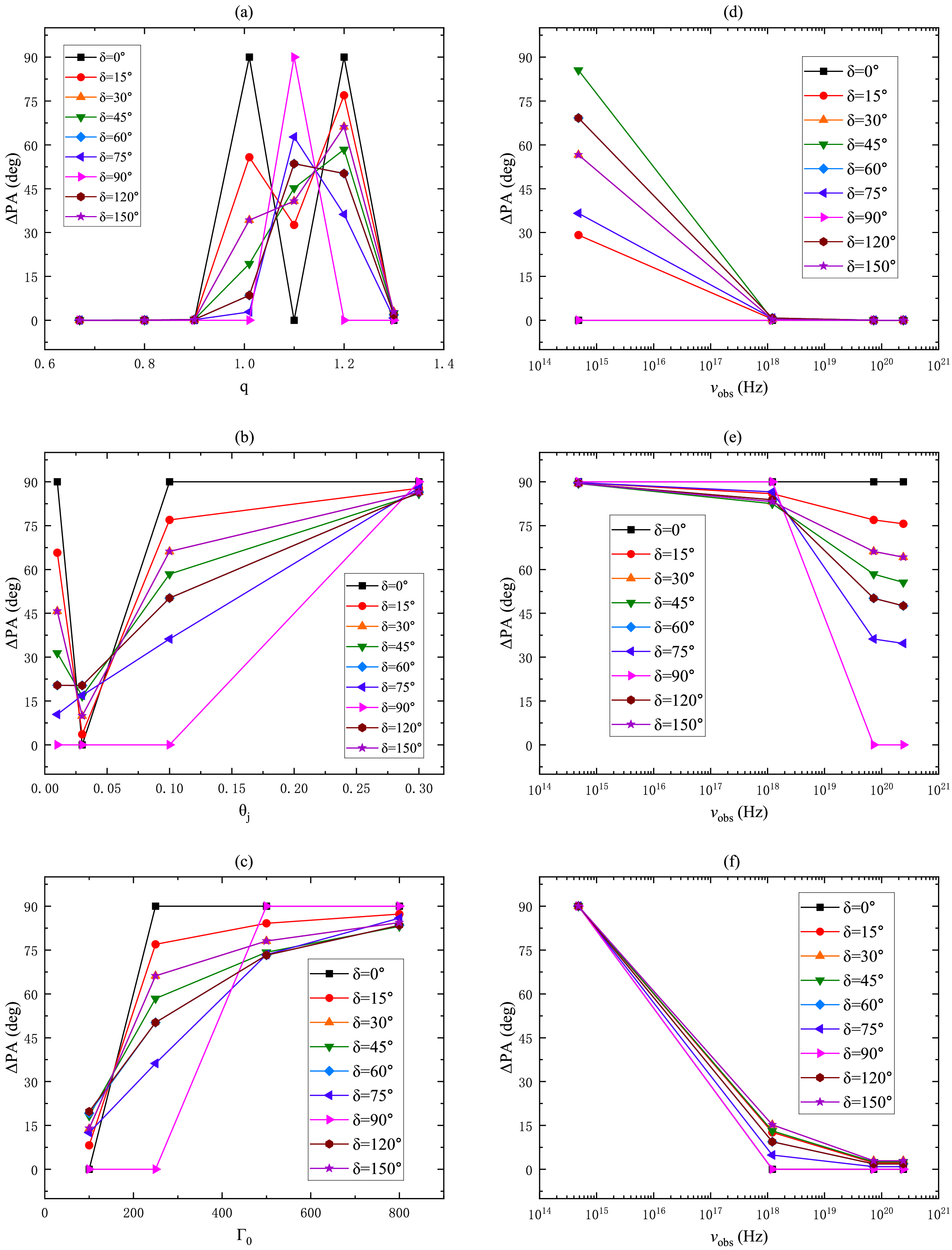}
\caption{The variations of $\Delta$PA with key parameters and the PA rotation spectra for various $\delta$ values. The (a), (b), (c) panels correspond to the variations of $\Delta$PA with $q$, $\theta_j$ and $\Gamma_0$, respectively. And the (d), (e), (f) panels correspond to the $\Delta$PA spectra with $q=2/3,\,1.2$ and $1.3$, respectively. In each panel, the different colored lines correspond to different $\delta$ values.}
\label{PA}
\end{figure*}

In our previous work, we found that three parameters: the bulk Lorentz factor $\Gamma_0$, the jet half-opening angle $\theta_j$, and $q=\theta_V/\theta_j$, where $\theta_V$ is the observation angle, have significant impacts on the PA rotation \citep{2024ApJ...973....2L}. The effects of these three key parameters on the PA rotations were investigated \citep{2025ApJ...979..219L}. Here, we discuss the $\Delta$PA spectrum and the variations of $\Delta$PA with the three key parameters at different magnetic field orientations. In Figure \ref{PA}, we show the variations of $\Delta$PA for various $\delta$ values with key parameters and the $\Delta$PA spectra. The $q-\Delta$PA, $\Gamma_0-\Delta$PA and $\theta_j-\Delta$PA curves are calculated at an observational energy of $300$ keV. The main reason for this choice is that the Gamma-Ray polarimeter, POLAR operated within the 50-500 keV energy band, and the forthcoming High-energy Polarimetry Detector (HPD) on board POLAR-2 will work in the energy bands of 30-800 keV. And 300 keV is roughly the midpoint of these energy ranges. And the $\Gamma_0-\Delta$PA and $\theta_j-\Delta$PA curves are calculated at an observational angle of $q=1.2$.

When $\delta<90^\circ$, for the $q-\Delta$PA curves (Figure \ref{PA} $(a)$) we can find that PA rotations are concentrated at slightly off-axis observations of $1\leq q\leq1.2$, independent of $\delta$. In the case of $1\leq q\leq1.2$, $\Delta$PA decreases and then increases with $q$ when $\delta=0^\circ$ and $15^\circ$. When $\delta$ is $30^\circ$ and $45^\circ$, $\Delta$PA increases with $q$. And when $\delta\geq60^\circ$, $\Delta$PA first increases and then decreases with $q$. The $\Gamma_0-\Delta$PA curves (Figure \ref{PA} $(c)$) show a general pattern that $\Delta$PA gradually increases with $\Gamma_0$. For $\Gamma_0\geq250$, with a few exceptions (such as $\delta=90^\circ$ for $\Gamma_0=500$), $\Delta$PA shows an increasing tendency with decreasing of $\delta$. For the $\theta_j-\Delta$PA curves (Figure \ref{PA} $(b)$), the typical value of the GRB jet half-opening angle is 0.1 rad \citep{2019MNRAS.488.5823L, 2020ApJ...893...77W, 2023ApJ...959...13R}. For the fitting of the energetic GRB, a narrow jet with $\theta_j\sim$0.01 rad is indicated \citep{2024JHEAp..41...42Z}. And the cases with $\theta_j$=0.03 and 0.3 rad are also included. $\Delta$PA shows an increasing tendency with $\theta_j$, except for the special case of $\theta_j=0.03$ rad. In the $\Delta$PA spectrum, $\Delta$PA has a tendency to decrease with the observational frequency, which in general agrees with our previous conclusions \citep{2025ApJ...979..219L}. For the energy band used for calculation, we take R-Band (optical), 5 keV (X-ray), 300 keV (gamma-ray), and 1 MeV (gamma-ray), of which we currently have the facilities or will have in the near future. And for the X-ray and gamma-ray band, the frequency within the energy range of the corresponding detector are used. In optical R band, we have Liverpool Telescope and Very Large Telescope. In X-ray band, we have IXPE (2-8 keV) and will have LPD (2-10 keV) on board POLAR-2. In gamma-ray band, we will have HPD (30-800 keV) on board POLAR-2. Because the polarization prediction of the photosphere model and the synchrotron model would differ significantly above the MeV energy band \citep{2018ApJ...856..145L, 2022ApJ...926..104P, 2024ApJ...970...10L}, the $\Delta$PA spectra were also calculated up to 1 MeV. For on-axis observations with $q=2/3$ (Figure \ref{PA} $(d)$), PA rotations occur only in the optical band, and $\Delta$PA has a tendency to increase and then decrease with $\delta$. In contrast, there is no rotation in both the X-ray band and gamma-ray bands. For slight off-axis observations with $q=1.2$ (Figure \ref{PA} $(e)$), the $\Delta$PA would decrease gradually with $\delta$ in gamma-ray bands, while the rotation of PA is minimized at $\delta = 45^{\circ}$ in the X-ray bands. And in the optical band, the value of $\Delta$PA is $90^\circ$ in each of the $\delta$ cases we chose. When $q=1.3$ (Figure \ref{PA} $(f)$), the $\Delta$PA values in the optical band are also concentrated at $90^\circ$, while there is almost no rotation in gamma-ray bands. In X-ray band, $\Delta$PA increases and then decreases with $\delta$. When $\delta>90^\circ$, in each panel of Figure \ref{PA} the values of $\Delta$PA at $\delta=120^\circ$ ($\delta=150^\circ$) are the same as those at $\delta=60^\circ$ ($\delta=30^\circ$). Therefore, when two magnetic field orientations are complementary angles, their $\Delta$PA values would be the same.

When the product of $\theta_j$ and $\Gamma_0$ is a constant value, different combinations of ($\theta_j$, $\Gamma_0$) have little effect on the $\Delta$PA value \citep{2024ApJ...973....2L}. The product value $\theta_j\Gamma_0$ represents the ratio between the jet half-opening angle and the $1/\Gamma$ cone. Here, we first test that under different aligned field orientations, whether or not $\Delta$PA value remains related only to the product value of $\theta_j\Gamma_0$ and not to the specific combinations of ($\theta_j$, $\Gamma_0$). We take $q=1.2$ and show the results in Table \ref{table q=1.2}. The observational frequency is set as $300$ keV. Except for $(\theta_j\Gamma_0,\,\delta)=(5,\,30^\circ)$, $(5,\,45^\circ)$ and $(5,\,60^\circ)$, where the difference in $\Delta$PA for different combinations will be slightly greater than $10^\circ$, most of the differences in $\Delta$PA for different combinations with the same product remain within $10^\circ$. Therefore, the choice of different field orientations would in general do not affect the above conclusions.

\begin{deluxetable*}{lccccccc}
\tabletypesize{\large}
\tablewidth{60pt}
\tablecaption{The Values of $\Delta$PA for Different Combinations of ($\theta_j$, $\Gamma_0$) with a Fixed Product Value of $\theta_j\Gamma_0$ for Different Field Orientations.\label{table q=1.2}}
\tablehead{
{$\theta_j\Gamma_0=5$}
&\colhead{\textcolor{black}{$\delta=0^\circ$}}
&\colhead{\textcolor{black}{$\delta=15^\circ$}}
&\colhead{\textcolor{black}{$\delta=30^\circ$}}
&\colhead{\textcolor{black}{$\delta=45^\circ$}}
&\colhead{\textcolor{black}{$\delta=60^\circ$}}
&\colhead{\textcolor{black}{$\delta=75^\circ$}}
&\colhead{\textcolor{black}{$\delta=90^\circ$}}
}
\startdata
 $(\theta_j,\Gamma_0)=(0.05,100)$ & 90 & 50.04 & 26.15 & 11.77 & 3.91 & 1.97 & 0 \\
 $(\theta_j,\Gamma_0)=(0.02,250)$ & 90 & 47.83 & 25.85 & 10.28 & 4.04 & 5.4 & 0 \\
 $(\theta_j,\Gamma_0)=(0.01,500)$ & 90 & 47.62 & 36.47 & 25.89 & 16.49 & 8.96 & 0 \\
\hline
 $\theta_j\Gamma_0=8$ & $\delta=0^\circ$ & $\delta=15^\circ$ & $\delta=30^\circ$ & $\delta=45^\circ$ & $\delta=60^\circ$ & $\delta=75^\circ$ & $\delta=90^\circ$ \\
\hline
 $(\theta_j,\Gamma_0)=(0.04,200)$ & 0 & 6.24 & 7.28 & 13.04 & 16.08 & 11.92 & 0 \\
 $(\theta_j,\Gamma_0)=(0.02,400)$ & 0 & 4.59 & 11.54 & 18.71 & 22.73 & 19.39 & 0 \\
 $(\theta_j,\Gamma_0)=(0.01,800)$ & 0 & 11.03 & 15.72 & 16.78 & 15.16 & 10.67 & 0 \\
\hline
 $\theta_j\Gamma_0=10$ & $\delta=0^\circ$ & $\delta=15^\circ$ & $\delta=30^\circ$ & $\delta=45^\circ$ & $\delta=60^\circ$ & $\delta=75^\circ$ & $\delta=90^\circ$ \\
\hline
 $(\theta_j,\Gamma_0)=(0.125,80)$ & 0 & 8.19 & 13.96 & 16.64 & 15.63 & 10.35 & 0 \\
 $(\theta_j,\Gamma_0)=(0.1,100)$ & 0 & 8.21 & 14.02 & 16.65 & 15.79 & 10.89 & 0 \\
 $(\theta_j,\Gamma_0)=(0.05,200)$ & 0 & 9.25 & 17.36 & 22.44 & 22.09 & 15.48 & 0 \\
\hline
 $\theta_j\Gamma_0=25$ & $\delta=0^\circ$ & $\delta=15^\circ$ & $\delta=30^\circ$ & $\delta=45^\circ$ & $\delta=60^\circ$ & $\delta=75^\circ$ & $\delta=90^\circ$ \\
\hline
 $(\theta_j,\Gamma_0)=(0.2,125)$ & 90 & 75.86 & 64.76 & 56.63 & 49.55 & 38.62 & 0 \\
 $(\theta_j,\Gamma_0)=(0.1,250)$ & 90 & 75.62 & 64.19 & 55.55 & 47.51 & 34.67 & 0 \\
 $(\theta_j,\Gamma_0)=(0.05,500)$ & 90 & 78.78 & 68.95 & 60.81 & 53.05 & 41.42 & 0 \\
\hline
 $\theta_j\Gamma_0=50$ & $\delta=0^\circ$ & $\delta=15^\circ$ & $\delta=30^\circ$ & $\delta=45^\circ$ & $\delta=60^\circ$ & $\delta=75^\circ$ & $\delta=90^\circ$ \\
\hline
 $(\theta_j,\Gamma_0)=(0.2,250)$ & 90 & 84.11 & 79.15 & 75.99 & 73.05 & 77.42 & 90 \\
 $(\theta_j,\Gamma_0)=(0.1,500)$ & 90 & 84.16 & 78.09 & 71.88 & 73.2 & 73.4 & 90 \\
 $(\theta_j,\Gamma_0)=(0.0625,800)$ & 90 & 83.95 & 78.77 & 75.3 & 71.92 & 76.02 & 90 \\
\hline
 $\theta_j\Gamma_0=80$ & $\delta=0^\circ$ & $\delta=15^\circ$ & $\delta=30^\circ$ & $\delta=45^\circ$ & $\delta=60^\circ$ & $\delta=75^\circ$ & $\delta=90^\circ$ \\
\hline
 $(\theta_j,\Gamma_0)=(0.2,400)$ & 90 & 87.5 & 84.8 & 83.65 & 84.13 & 86.4 & 90 \\
 $(\theta_j,\Gamma_0)=(0.1,800)$ & 90 & 86.38 & 83.37 & 81.62 & 81.86 & 84.76 & 90 \\
 $(\theta_j,\Gamma_0)=(0.08,1000)$ & 90 & 86.97 & 84.52 & 83.21 & 83.59 & 85.01 & 90 \\
\hline
\enddata
\end{deluxetable*}

In order to further discuss the conclusions obtained in Figure \ref{PA} and investigate the PA rotation under different field orientations, we choose a number of specific $\delta$ values and calculate the distribution of the $\Delta$PA in the plane of ($q$, $\theta_j\Gamma_0$). In Figure \ref{deltaPA}, the distributions of $\Delta$PA (with $\Delta$PA $>10^\circ$) in the ($q$, $\theta_j\Gamma_0$) plane under various orientation are shown. The distribution laws of $\Delta$PA are in general independent of $\delta$ values. When the orientation $\delta$  equals to $0^\circ$ or $90^\circ$ as shown in Figure \ref{deltaPA} as $(e)$ and $(f)$ panels, there are only two possible values for $\Delta$PA in the range of $T_{90}$: $0^\circ$ and $90^\circ$. The range of $\Delta$PA distribution is from $q=1+1/(13\theta_j\Gamma_0)$ to $q=1.2+4/(\theta_j\Gamma_0)$, which is same as other orientations. When $\delta=90^\circ$, the $90^\circ$ PA rotations only occur in the region where $1<q<1.2$ and $\theta_j\Gamma_0>10$. In contrast, the distrbution range of $\Delta$PA in ($q$, $\theta_j\Gamma_0$) plane is much wider for $\delta=0^\circ$.

When the field orientations do not equal to $0^\circ$ and $90^\circ$, the distribution laws of $\Delta$PA in the ($q$, $\theta_j\Gamma_0$) plane are independent of the field orientations. The range of $q$ with $\Delta$PA $>10^\circ$ is still from $q=1+1/(13\theta_j\Gamma_0)$ to $q=1.2+4/(\theta_j\Gamma_0)$ for a fixed value of $\theta_j\Gamma_0$. As the $\theta_j\Gamma_0$ value increases, the range of $q$ for $\Delta$PA $>10^\circ$ becomes smaller. When $q<1.2$, for a fixed value of $q$, $\Delta$PA has a tendency to increase and then decrease and then increase again as $\theta_j\Gamma_0$ increases. Large PA rotations would occur in the regions of $1<q<1.2$ and $\theta_j\Gamma_0>50$, which generally agrees with the results obtained in \cite{2024ApJ...973....2L}. When $q>1.2$, for a fixed value of $q$, the $\Delta$PA increases and then decreases with $\theta_j\Gamma_0$. For a fixed value of $\theta_j\Gamma_0$, $\Delta$PA increases with $q$ when $\theta_j\Gamma_0\leqslant 1$. These patterns are also consistent with the calculated case of $\delta=30^\circ$ in our previous paper \citep{2024ApJ...973....2L}. The distribution range of the PA rotations with $\Delta$PA $>10^\circ$ in ($q$, $\theta_j\Gamma_0$) plane decreases gradually with increasing $\delta$, while the $\Delta$PA distribution laws in the plane were roughly hold. When $\delta=75^\circ$, there are a few points in ($q$, $\theta_j\Gamma_0$) plane with $\Delta$PA $>10^\circ$ in the area of $\theta_j\Gamma_0<10$.

The distributions of $\Delta$PA (with $\Delta$PA $>10^\circ$) in the ($q$, $\theta_j\Gamma_0$) plane under orientations of $\delta=120^\circ$ and $150^\circ$ are shown in Figure \ref{deltaPA} as $(g)$ and $(h)$ panels. Consistent with the findings in Figure \ref{PA}, when the angles of the two magnetic field orientations are complementary, i.e. $\delta=120^\circ$ and $\delta=60^\circ$, as well as $\delta=150^\circ$ and $\delta=30^\circ$, the $\Delta$PA values are the same under identical parameters.

\begin{figure*}
\centering
\includegraphics[width=0.5\textwidth]{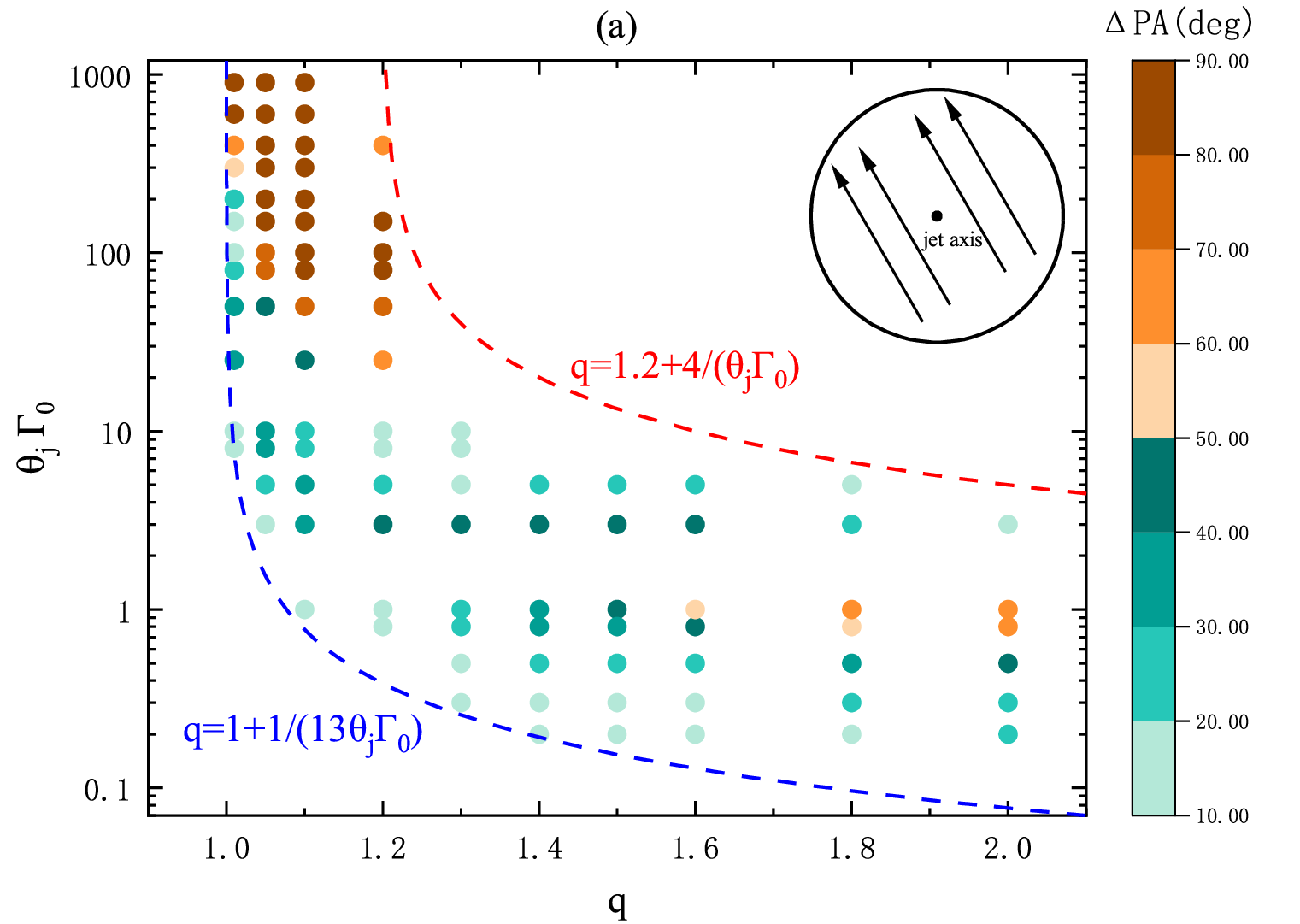}\includegraphics[width=0.5\textwidth]{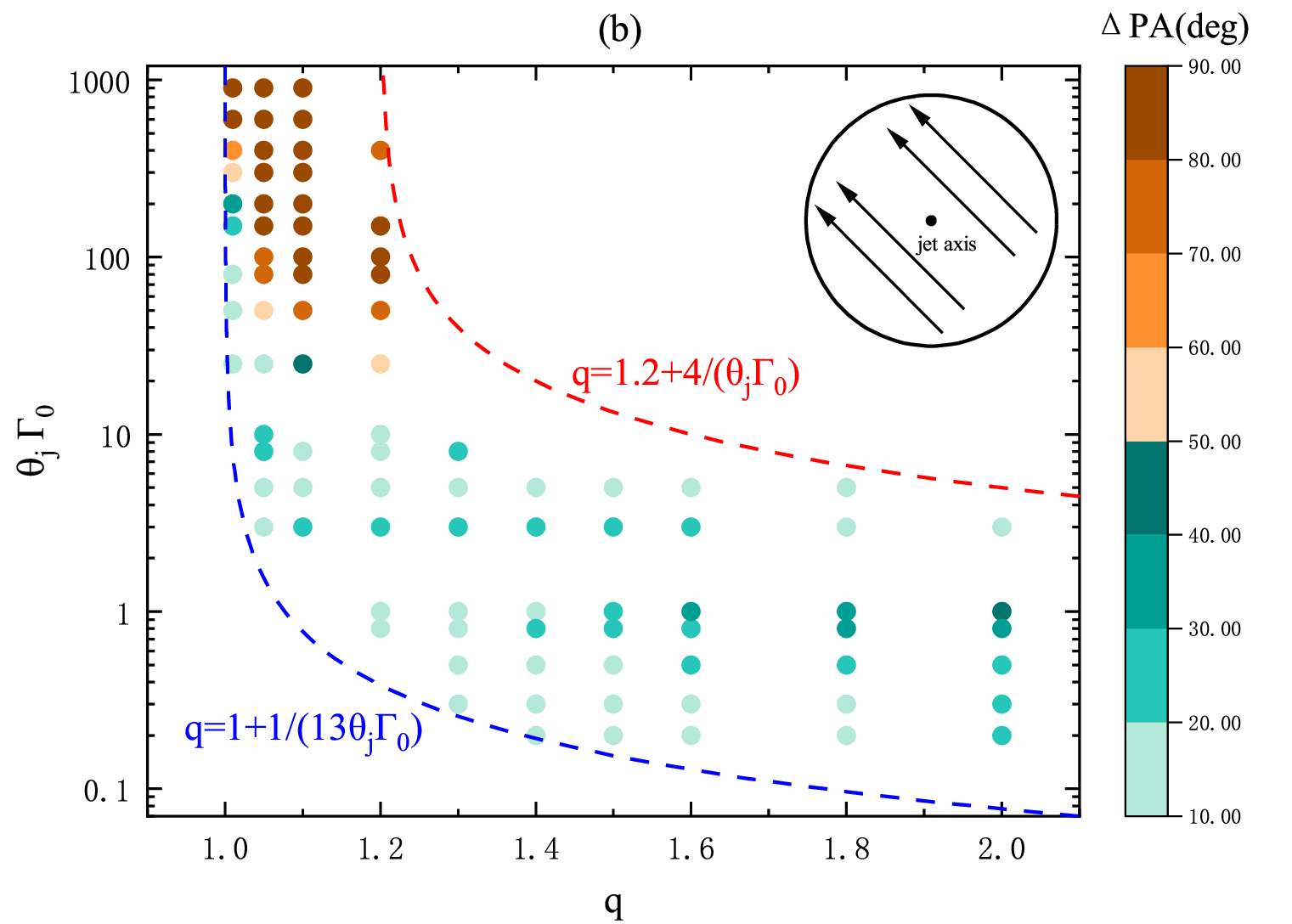}
\includegraphics[width=0.5\textwidth]{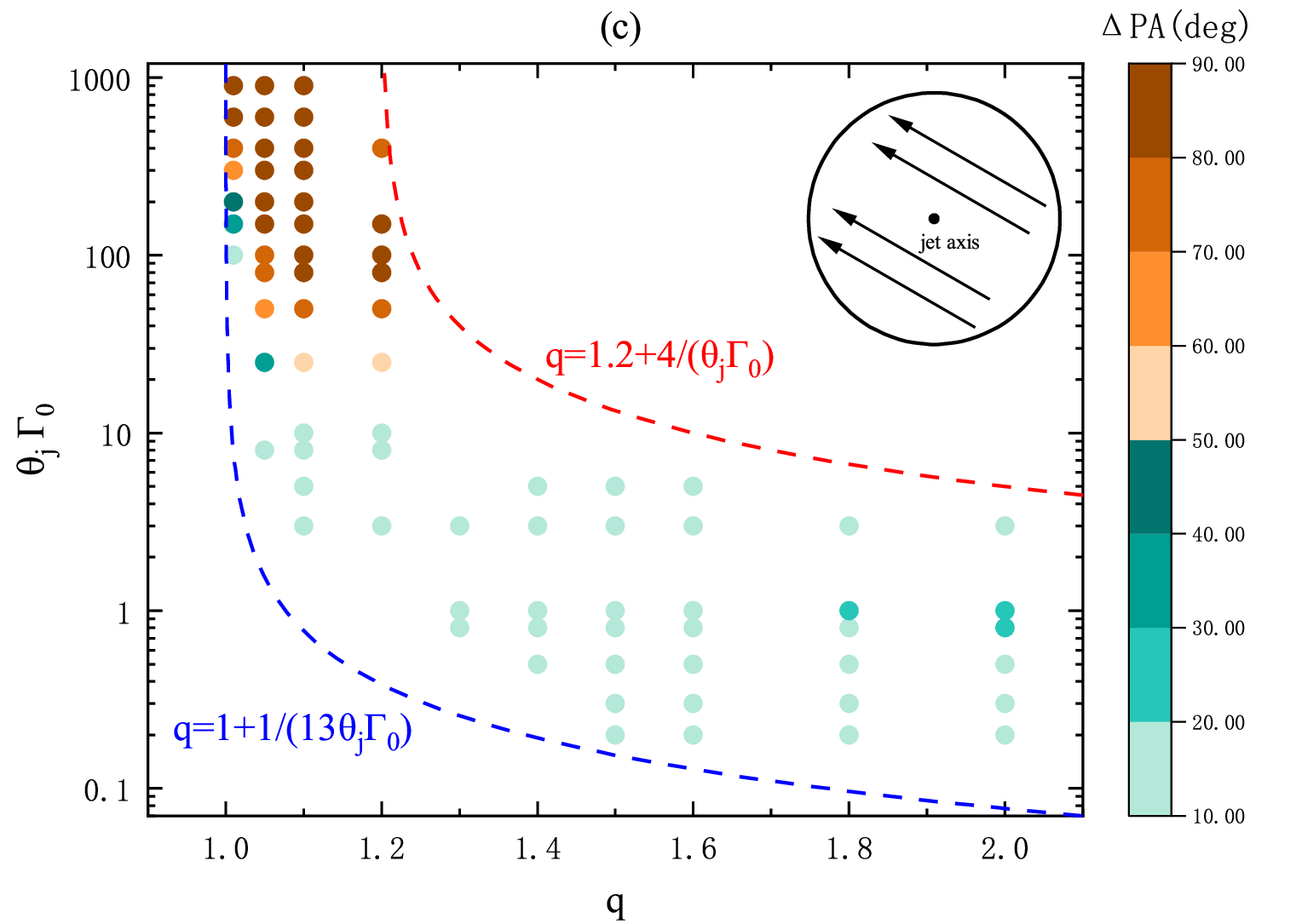}\includegraphics[width=0.5\textwidth]{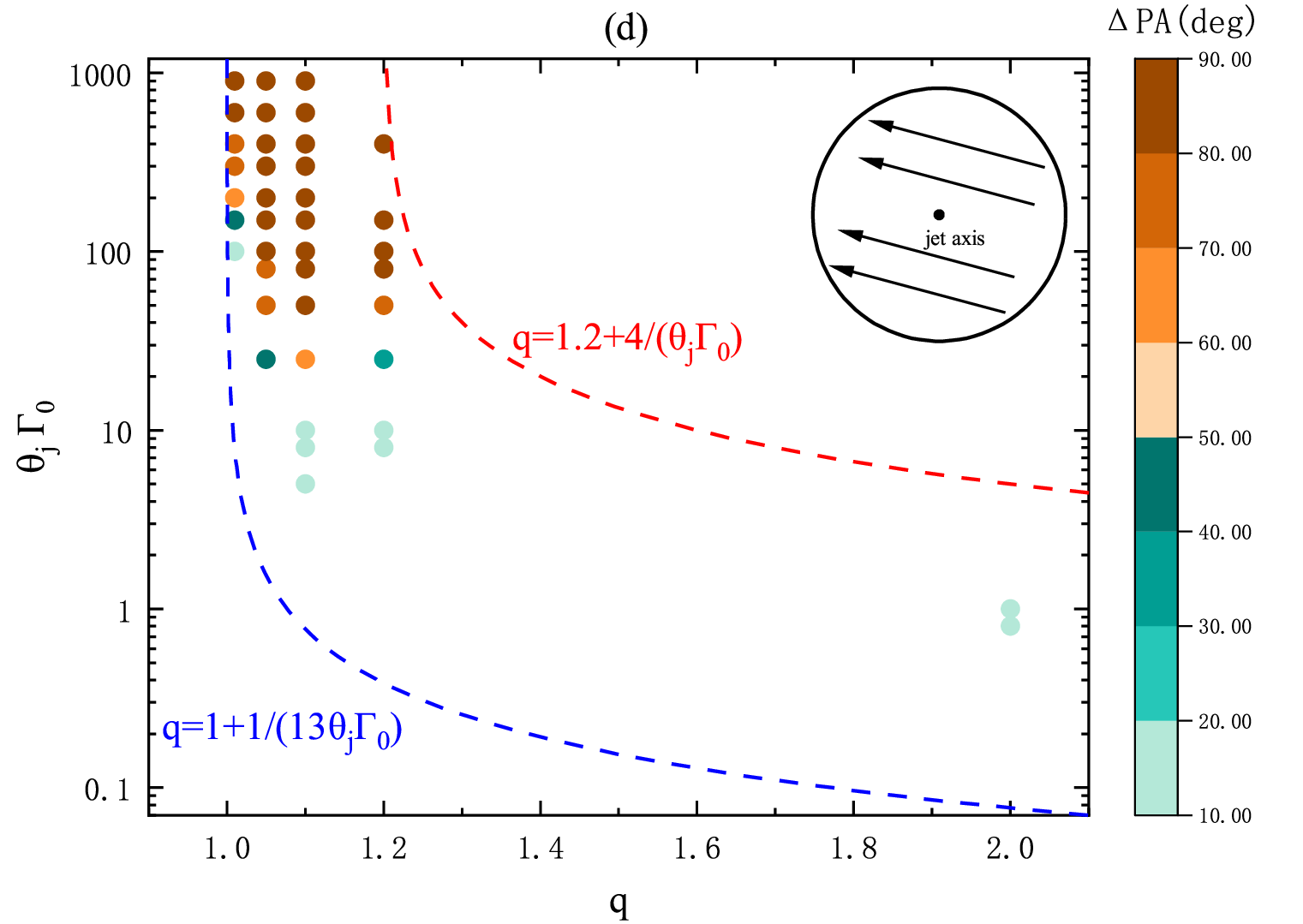}
\includegraphics[width=0.5\textwidth]{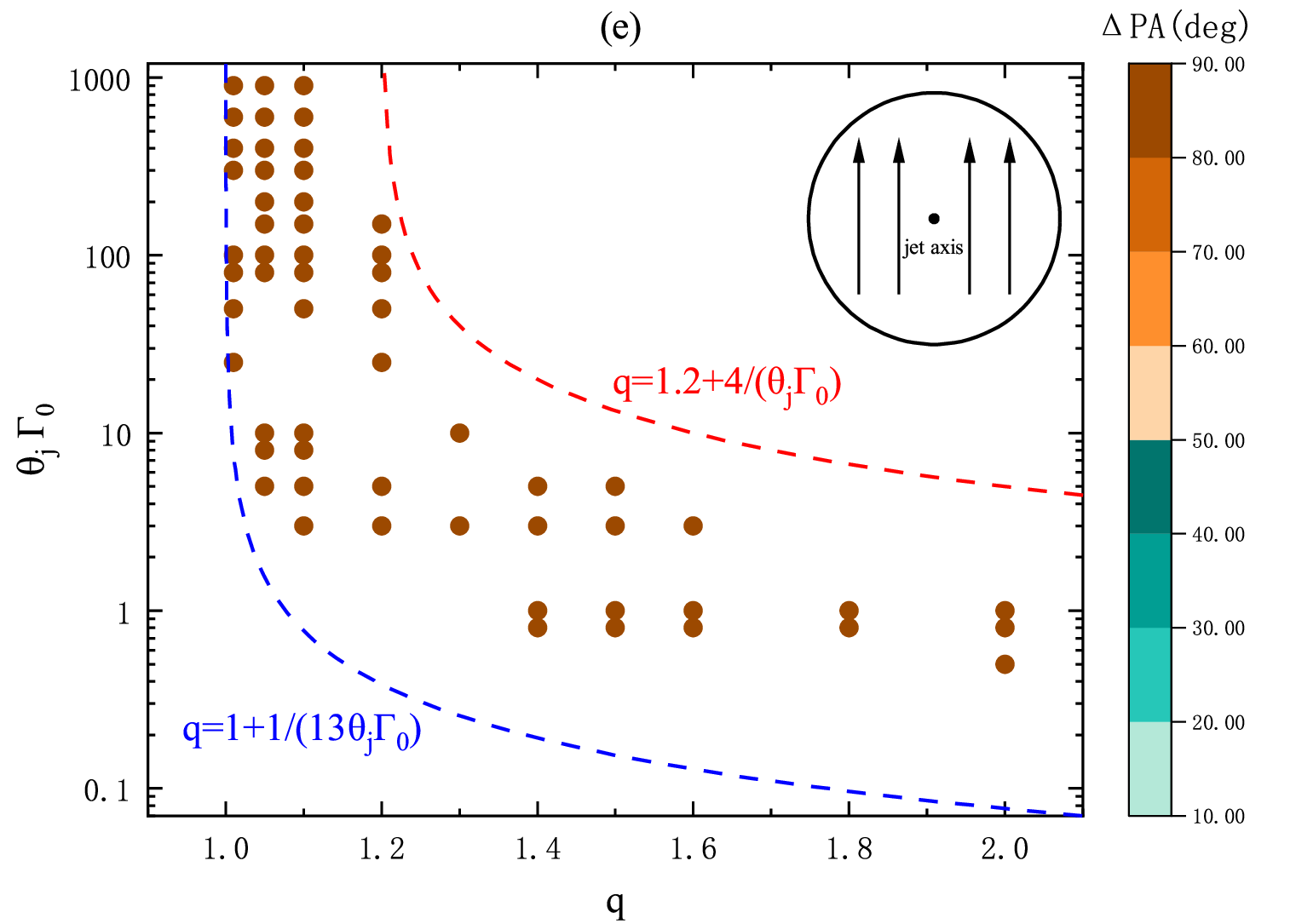}\includegraphics[width=0.5\textwidth]{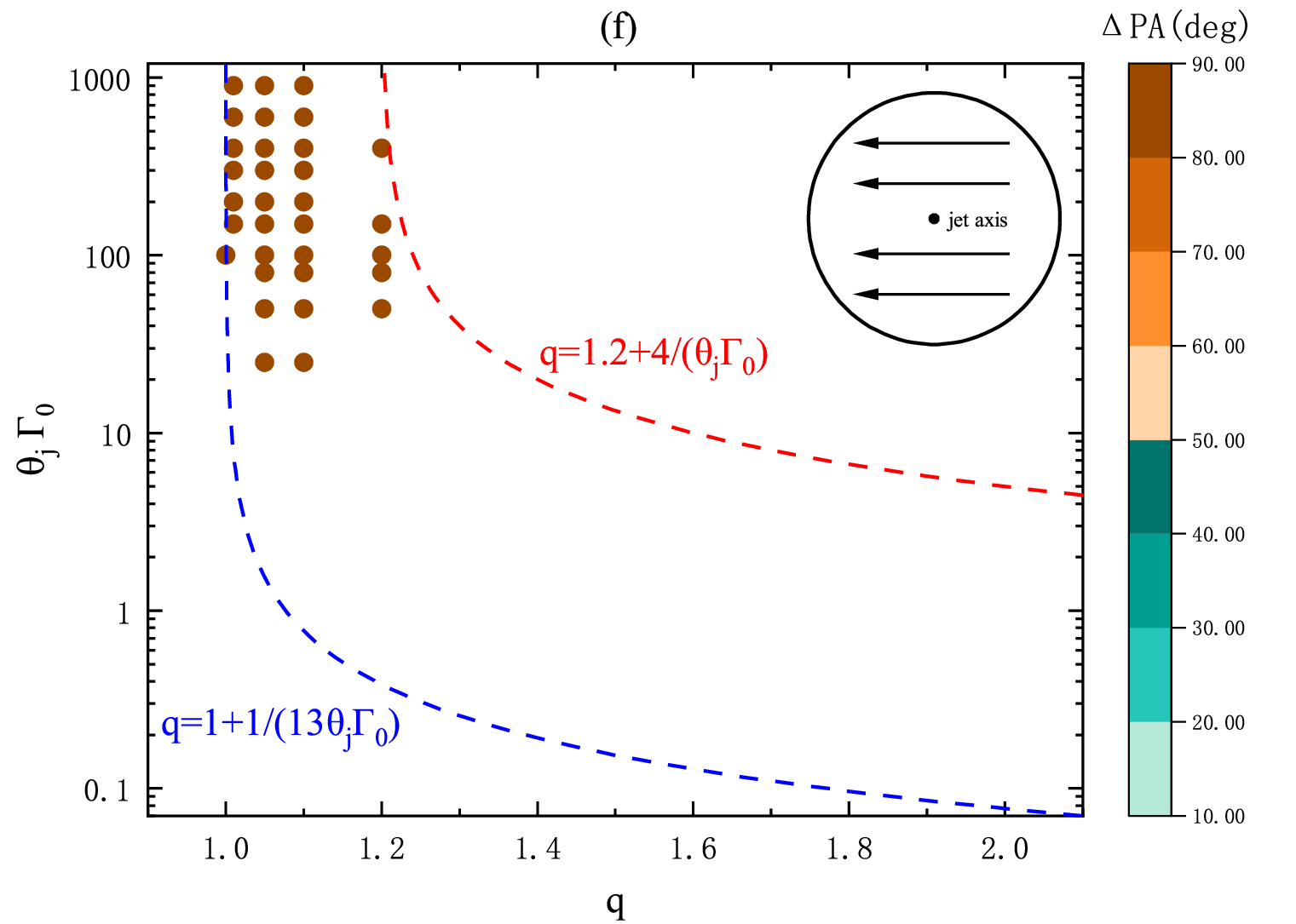}
\caption{Distribution of the $\Delta$PA in the ($q$, $\theta_j\Gamma_0$) plane. The panels (a), (b), (c), (d), (e), (f), (g), (h) correspond to the field orientations ($\delta$) of $30^\circ$, $45^\circ$, $60^\circ$, $75^\circ$, $0^\circ$, $90^\circ$, $120^\circ$, and $150^\circ$, respectively. In each panel, the red-dashed and blue-dashed reference lines are the boundaries of the parameter space with $\Delta$PA $>10^\circ$. In the upper right corner of each panel, the orientation of the corresponding aligned field on the jet surface is shown. }
\label{deltaPA}
\end{figure*}
\begin{figure*}
\centering
\includegraphics[width=0.5\textwidth]{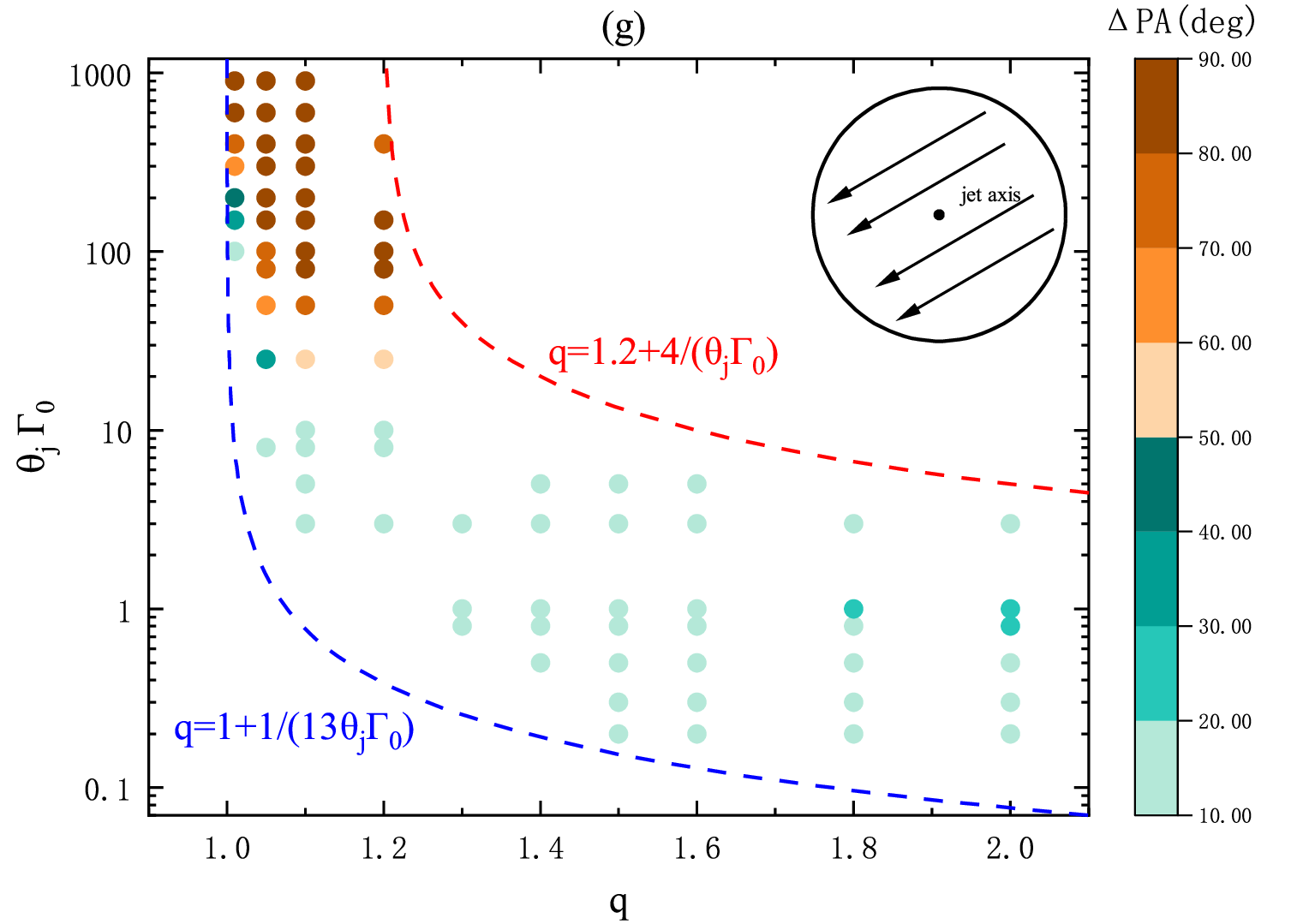}\includegraphics[width=0.5\textwidth]{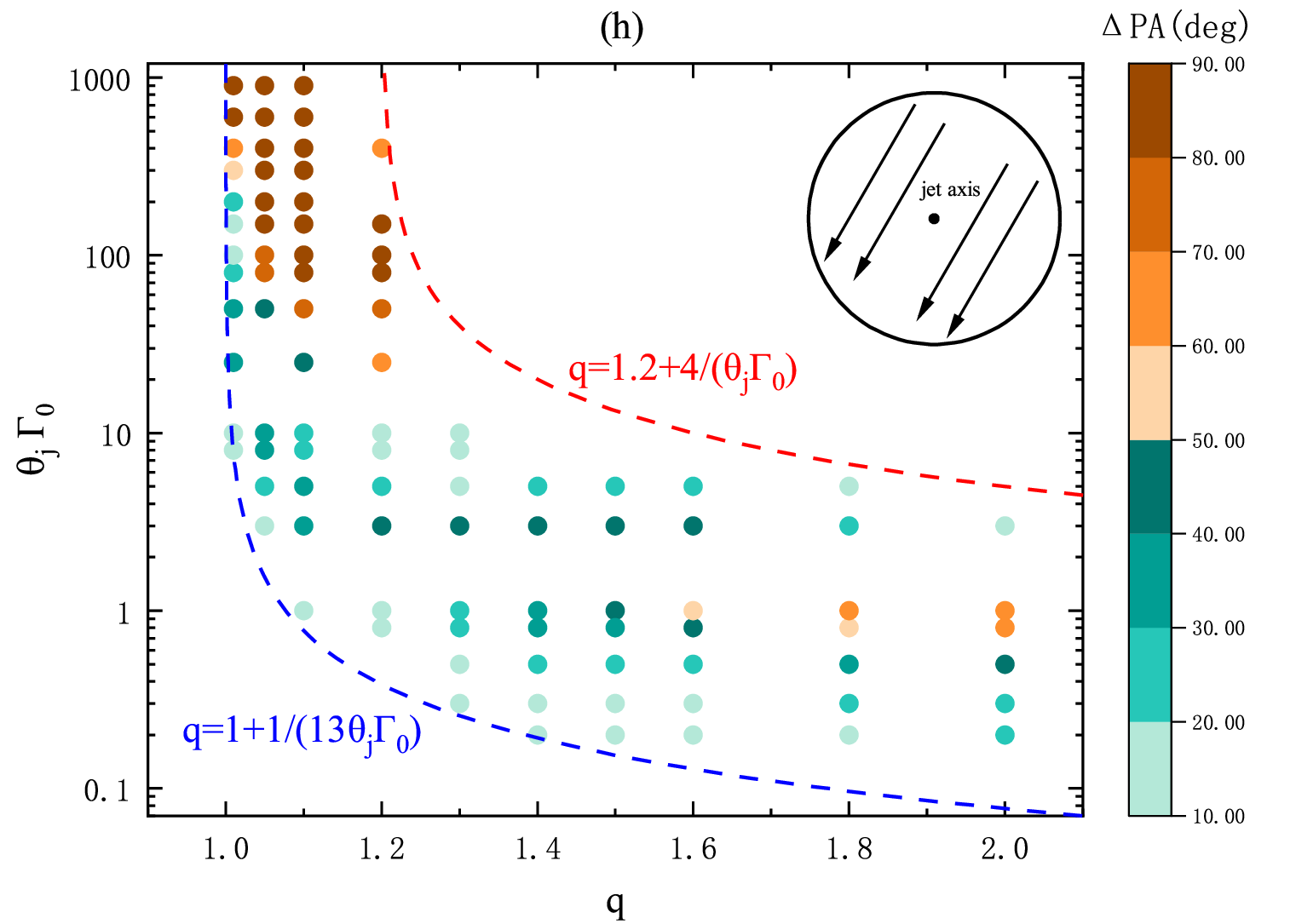}
\addtocounter{figure}{-1}
\caption{(Continued)}
\end{figure*}

\section{Conclusions and discussion}\label{sec:Conclusions and Discussion}

In this paper, we use a simple parametric magnetic reconnection model to study the PA rotation during the GRB prompt phase, especially the impact of the field orientation on the PA rotations. The laws found in our previous paper \citep{2023ApJ...946...12W, 2024ApJ...973....2L, 2025ApJ...979..219L} under the field orientation $\delta=30^\circ$ would in general hold under other orientations. The $\Delta$PA values would be the same for two complementary field orientations.

In the $\Delta$PA spectrum, we find that $\Delta$PA has a tendency to decrease with the observational frequency. The maximum PA rotation in each energy band occurs for slightly off-axis, and significant PA rotation for on-axis observations only occurs in optical band. These conclusions are independent of the field orientation and consistent with our previous studies at an orientation of $\delta=30^\circ$ \citep{2023ApJ...946...12W, 2025ApJ...979..219L}.

The distributions of $\Delta$PA in ($q$, $\theta_j\Gamma_0$) plane for different field orientations were also considered. The conclusions of the previous work that PA rotations would only depend on the product value of $\theta_j\Gamma_0$ when $\delta=30^\circ$ and the $q$ ranges for $\Delta$PA $>10^\circ$ become narrow as the value of $\theta_j\Gamma_0$ increases \citep{2024ApJ...973....2L}, still hold for other orientations. Generally, the rotations with $\Delta$PA $>10^\circ$ will happen within the region from $q=1+1/(13\theta_j\Gamma_0)$ to $q=1.2+4/(\theta_j\Gamma_0)$ for a fixed value of $\theta_j\Gamma_0$. When the value of the magnetic field direction $\delta$ becomes larger, the distribution region of $\Delta$PA with $\Delta$PA $>10^\circ$ in ($q$, $\theta_j\Gamma_0$) plane will gradually become smaller and eventually converges to the region of $1<q<1.2$ and $\theta_j\Gamma_0>20$ for $\delta=90^\circ$. For all the field orientations, consistent with previous studies with $\delta=30^\circ$ \citep{2024ApJ...973....2L}, large PA rotations with $\Delta$PA$=90^\circ$ would occur in the regions of $1<q<1.2$ and $\theta_j\Gamma_0>50$.

Additionally, we find two specific field orientations $\delta=0^\circ$ and $90^\circ$. For these two orientations, there will only be $0^\circ$ and $90^\circ$ rotations of the PA within $T_{90}$. Therefore, if a non $90^\circ$ PA rotation is observed during the prompt phase, the orientation of the magnetic field in the emission region cannot be an aligned magnetic field with $\delta=0^\circ$ or $90^\circ$. The reasons for PA rotations are complex and had been roughly explored in the previous papers \citep{2023ApJ...946...12W,2024ApJ...973....2L,2024A&A...687A.128C,2025ApJ...979..219L}. However, they are crucial. There are still degeneracies in parameters in case that PD curve and PD spectrum are considered. And PA curve, PA spectrum and $\Delta$PA spectrum may break these degeneracies.

\begin{acknowledgements}
This work is supported by the National Natural Science Foundation of China (grant No. 12473040). M.X.L would also like to acknowledge the financial support from Jilin University.
\end{acknowledgements}


\listofchanges

\bibliography{ref}

\end{document}